\documentclass[a4paper,11pt,DIV12]{scrartcl}

\addtokomafont{disposition}{\rmfamily}

\usepackage[utf8]{inputenc}
\usepackage{enumitem}
\usepackage{amsmath}
\usepackage{amssymb}
\usepackage{latexsym}
\usepackage{microtype}
\usepackage{xspace}
\usepackage{hyperref}
\usepackage{xcolor}
\usepackage{tikz}
\usetikzlibrary{patterns,snakes,arrows}
\usepackage{ifthen}
\usepackage{orcidlink}

\newenvironment{mk}{\noindent\color{red}} {}

\newtheorem{theorem}{Theorem}

\newtheorem{corollary}[theorem]{Corollary}

\newenvironment{proof}[1][]{\pagebreak[3]\noindent\textit{Proof\ifthenelse{\equal{#1}{}}{}{ (#1)}: }}{\pagebreak[3]\medskip}

\setlist{noitemsep}

\newcommand{\N}{\mathbb{N}}

\newcommand{\set}[2]{\left\{\hspace*{0.5pt}#1\vphantom{#2}\hspace*{0.5pt} \left|\hspace*{2pt} \vphantom{#1}#2 \right.\hspace*{0.5pt}\right\}}
\newcommand{\oneset}[1]{\left\{\hspace*{0.5pt}#1\hspace*{0.5pt}\right\}}
\newcommand{\smallset}[1]{\left\{#1\right\}}

\newcommand{\abs}[1]{\left|\mathinner{#1}\right|}

\makeatletter
\newcommand\klammeraffe{@}
\makeatother

\newcommand{\qed}{\hspace*{\fill}$\Box$}

\newcommand{\RA}{\Rightarrow}
\newcommand{\RAS}{\stackrel{*}{\RA}}
\newcommand{\ms}{\hspace*{1pt}} % mini space

\newcommand{\coloneq}{\mathrel{\mathop:}=}

\title{Yet another proof of Parikh’s Theorem}

\author{Manfred Kuf\-leitner\hspace*{2pt}\orcidlink{0000-0003-3869-416X}
}

\date{\normalsize{University of Stuttgart, FMI, Germany\\
\texttt{kufleitner{\klammeraffe}fmi.uni-stuttgart.de}}}

\begin{document}

\makeatother
\maketitle

\begin{abstract}
\noindent
\textbf{Abstract.}\ 
Parikh's Theorem says that the Parikh image of a context-free language is semilinear.
We give a short proof of Parikh's Theorem using the formulation of Verma, Seidl, and Schwentick in terms of Presburger arithmetic. The proof relies on an Eulerian property of derivation trees of context-free languages and was inspired by Hierholzer's algorithm; it does not use the Chomsky normal form.
\end{abstract}

\noindent
These notes are based on a presentation given in Bordeaux on June 16th, 2015 at a workshop in honor of Volker Diekert's 60th birthday.

A \emph{context-free grammar} $G=(N,\Sigma,P,S)$ consist of a finite set of \emph{non-terminals} $N$, a finite alphabet $\Sigma$, a finite set of \emph{productions} $P \subseteq N \times (N \cup \Sigma)^*$, and a \emph{start symbol} $S$. The elements of $\Sigma$ are also called \emph{terminals} and the productions are also known as \emph{rules}. The rules are usually written as $X \to u$ for $X \in N$ and $u \in (N \cup \Sigma)^*$; they induce a string rewrite system on $(N \cup \Sigma)^*$: For $v,w \in (N \cup \Sigma)^*$ we write $v \RA w$ if $v = pXq$ and $w = puq$ for $p,q \in (N \cup \Sigma)^*$ and a rule $X \to u$ in~$P$. We consider the reflexive and transitive closure $\RAS$ of the relation $\RA$. If $v \RAS w$, then $w$ is a \emph{derivation} of $v$. The language $L(G)$ generated by $G$ consists of all derivations of the start symbol $S$ which do not contain any non-terminal:
\begin{equation*}
  L(G) = \set{w \in \Sigma^*}{S \RAS w}
\end{equation*}
A language is \emph{context-free} if it is generated by a context-free grammar.

A \emph{derivation tree} is an ordered tree $t$ with labeled nodes. The label of the root is the start-symbol $S$, the labels for the inner nodes come from the non-terminals $N$, all leafs have a label in $\Sigma \cup\smallset{\varepsilon}$ where $\varepsilon$ denotes the empty word, and if an inner node has label $X$, then the sequence of labels of its children forms a word $u \in (N \cup \Sigma)^*$ such that $X \to u$ is in $P$. If the sequence of labels of the leafs forms a word $w$, then we say that $t$ is a derivation tree for $w$. We have $S \RAS w$ if and only if there exists a derivation tree for $w$. There can be more than one derivation tree for $w$.

Let $\abs{w}_a$ denote the number of occurrences of the letter $a$ in the word $w \in \Sigma^*$. The \emph{Parikh image} of $w$ is 
\begin{equation*}
  \Psi(w) = \left(\abs{w}_a\right)_{a \in \Sigma}
\end{equation*}
As usual, for $L \subseteq \Sigma^*$ we set $\Psi(L) = \set{\Psi(w)}{w\in L}$.
If $\Sigma = \oneset{a_1,\ldots,a_k}$, then the Parikh mapping $\Psi : \Sigma^* \to \N^{k}$ is the homomorphism defined by $\Psi(a_i) = (0,\ldots,0,1,0,\ldots,0)$ with a one at the $i$-th component; here, the operation on $\N^k$ is component-wise addition.

A subset of the form $\set{u + x_1 v_1 + \cdots + x_n v_n}{x_i \in \N} \subseteq \N^k$ for $u,v_i \in \N^k$ is called \emph{linear}. A \emph{semilinear} set is a finite union of linear sets. Parikh's Theorem says that Parikh images of context-free languages are semilinear. 
%The converse is not true; for instance the Parikh image of the non-context-free language $L=\set{a^n b^n c^n}{n \geq 0}$ is the linear set $\set{(n,n,n) \in \N^3}{n \geq 0} = \set{x \cdot (1,1,1)}{x \in \N}$, but $L$ is not context-free. 
We equivalently state Parikh's Theorem in terms of Presburger arithmetic, the first-order logic of $(\N,+)$. A formula in Presburger arithmetic is \emph{existential} if there is no universal quantifier and no negation.
We use the proof of Verma, Seidl, and Schwentick~\cite{VermaEtAl2005cade} to construct a Presburger formula for a given context-free grammer $G$. While~\cite{VermaEtAl2005cade} relies on a result of Esparza~\cite{Esparza1997fi} for showing that every model of the formula has a corresponding word in $L(G)$, we give a new self-contained proof for this part.

\begin{theorem}[Verma, Seidl, Schwentick~\cite{VermaEtAl2005cade}]\label{thm:Parikh}
Given a context-free grammar $G$, one can compute an existential Presburger formula for the Parikh image of $L(G)$ in linear time.
\end{theorem}

\begin{proof}
Let $G = (N,\Sigma,P,S)$ with $\abs{\Sigma} = k$. For a rule $p = (Y \to u)$ and $X \in N \cup \Sigma $ we define
\begin{equation*}
  X(p) = \begin{cases}
  1 & \text{ if } X = Y \\
  0 & \text{ otherwise}
  \end{cases}
\end{equation*}
and we set $\abs{p}_X = \abs{u}_X$. We define a Presburger formula $\varphi$ for $\Psi\big(L(G)\big)$. For every rule $p \in P$ we introduce a variable $x_p$ which represents the number of applications of $p$. For $X \in N \setminus \smallset{S}$ we define
\begin{equation*}
  \alpha_X \coloneq \Bigg( \sum_{p \in P} X(p) \cdot x_p = \sum_{p \in P} \abs{p}_X \cdot x_p \Bigg)
\end{equation*}
and
\begin{equation*}
  \alpha_S \coloneq \Bigg( \sum_{p \in P} S(p) \cdot x_p = 1 + \sum_{p \in P} \abs{p}_S  \cdot x_p \Bigg)
\end{equation*}
The formula $\alpha_X$ says that the number of occurrences of $X$ on the left side of a rule application matches the number of $X$'s on the right side of a rule application. Since $S$ is the start symbol, there is one more occurrence on the left than on the right. 
For every non-terminal $X \in N$ we introduce an index $y_X$ for numbering consecutively the first occurrences of the non-terminals in the derivation tree. We ensure that $y_X = 0$ if and only if $X$ does not occur. Let $\beta_S \coloneq (y_S = 1)$ and for $X \neq S$ we define
\begin{align*}
\beta_X \coloneq \,&\Bigg(y_X = 0 \,\wedge\, \sum_{p\in P} \abs{p}_X \cdot x_p = 0\Bigg) \vee 
\\ &\bigvee_{\substack{(Y \to u) \ms\in\ms P \\ \abs{u}_X > 0}} \!\!\!\big(x_{Y\to u} > 0 \,\wedge\, y_Y > 0 \,\wedge\, y_X = y_Y + 1\big)
\end{align*}
The formula $\varphi$ has a free variable $z_a$ for the number of occurrences of the letter $a \in \Sigma$. We define
\begin{equation*}
  \gamma_a \coloneq \Bigg( z_a = \sum_{p \in P} \abs{p}_a \cdot x_p \Bigg)
\end{equation*}
The formula $\varphi$ is the conjunction of all these formulas, with existentially quantified variables $x_p$ and $y_X$, i.e., for $\overline{x} = (x_p)_{p \in P}$ and $\overline{y} = (y_X)_{X \in N}$ we set
\begin{equation*}
\varphi \,\coloneq \; \exists \overline{x} \,\exists \overline{y} \colon \! \bigwedge_{X \in N} \!(\alpha_X \,\wedge\, \beta_X)  \,\wedge\, \!\bigwedge_{a \in \Sigma}\! \gamma_a 
\end{equation*}
If $w \in L(G)$, then there exists a derivation tree for $w$. Let $x_p$ be the number of occurrences of the rule $p \in P$ in this tree. Every inner node corresponds to both an occurrence on the left side of a rule and an occurrence on the right side of a rule, the sole exception is the root $S$. Thus $\overline{x} \models \bigwedge_{X \in N} \alpha_X$. We define $y_X$ inductively. Let $y_S = 1$ and if $y_X$ is not yet defined but for some parent $Y$ of $X$ the value $y_Y$ is already defined, then we set $y_X = y_Y + 1$.
Note that all non-terminals $X$ in the derivation tree satisfy $y_X > 0$ since there cannot be a topmost undefined non-terminal $X$ with a previously defined parent $Y$. All ultimately undefined variables $y_X$ are set to zero. Whenever we define $y_X > 0$ for $X \neq S$, then this is due to some rule $Y \to u$ with $\abs{u}_X > 0$ for some previously defined variable $y_Y$; in particular $x_{Y \to u} > 0$. Therefore $\overline{x},\overline{y} \models \bigwedge_{X \in N} \beta_X$. Let $z_a = \abs{w}_a$ and $\overline{z} = (z_a)_{a \in \Sigma}$. Every letter $a$ of $w$ is a leaf of the derivation tree, and every leaf in the derivation tree originates from some rule. This yields $\overline{x},\overline{z} \models \bigwedge_{a \in \Sigma} \gamma_a$. We conclude $\overline{z} \models \varphi$ and $\Psi\big(L(G)\big) \subseteq L(\varphi)$.

For the remaining inclusion $L(\varphi) \subseteq \Psi\big(L(G)\big)$ we  show that if $(z_a)_{a\in \Sigma} \models \varphi$, then there exists a derivation tree for a word $w$ with $\abs{w}_a = z_a$ for all $a \in \Sigma$.
Let $\overline{z} = (z_a)_{a\in \Sigma} \in \N^k$ with $\overline{z} \models \varphi$. Then there exist numbers $\overline{x} = (x_p)_{p\in P}$ and $\overline{y} = (y_X)_{X\in N}$ with 
\begin{equation*}
  \overline{x}, \overline{y}, \overline{z} \models \bigwedge_{X \in N} \!(\alpha_X \,\wedge\, \beta_X)  \,\wedge\, \!\bigwedge_{a \in \Sigma}\! \gamma_a 
\end{equation*}
We take $x_p$ copies of the rule $p$. If $p$ is of the form $X \to A_1 \cdots A_\ell$ with $A_i \in N \cup \Sigma$, then we represent $p$ as a tree:
\begin{center}
\hspace*{\fill}
\begin{tikzpicture}[level 1/.style={sibling distance=9mm,
  level distance=10mm}]
\node (X) {$X$}
  child {node {$A_1$}}
  child {node {$A_2$}}
  child {node (c) {$\cdots$}}
  child {node {$A_{\ell -1}$}}
  child {node {$A_{\ell}$}};
\draw (c) ++(0,-0.5) node[below] {tree for $\ell > 0$};
\end{tikzpicture}
\hspace*{\fill}
\begin{tikzpicture}[level 1/.style={sibling distance=9mm,
  level distance=10mm}]
\node (X) {$X$}
  child {node (c){$\varepsilon$}};
\draw (c) ++(0,-0.5) node[below] {tree for $\ell = 0$};
\end{tikzpicture}
\hspace*{\fill}
\end{center}
Let $r_X$ be the number of occurrences of $X$ as a root in all these trees, and let $s_X$ be the number of leafs with label $X$. The formula $\alpha_X$ ensures $r_X = s_X$ for $X \neq S$ and $r_S = s_S + 1$. The formula $\gamma_a$ says that there are $z_a$ leafs with label $a$. We iterate the following procedure as long as possible: If there is a tree $t_1$ with leaf $X$ and another tree $t_2$ with root $X$, the trees $t_1$ and $t_2$ are replaced with a new tree $t_1 \cdot t_2$ by identifying an $X$-leaf of $t_1$ with the root of $t_2$.
This reduces $r_X$ by $1$ and $s_X$ by $1$. Thus throughout we have $r_X = s_X$ for $X \neq S$ and $r_S = s_S + 1$. The number of leafs with label $a \in \Sigma$ remains unchanged.
Eventually, there cannot be two trees with root $X$ since then there has to be some tree with leaf $X$, which in turn could then be combined with one of the trees with root $X$ into a new tree. Thus we end up with a tree with root $S$ in which all leafs are terminals and with a set of trees of the following form: The root is $X \in N \setminus \smallset{S}$ and all leafs are terminals except for one leaf which is $X$.

Suppose that we have a tree $t$ with root $X$ and leaf $X$. If there exists another tree $s$ (possibly the tree with root $S$) with some node labeled by $X$, then we can factorize $s = p \cdot q$ at the non-terminal $X$. We replace the two trees $s$ and $t$ by the tree $p \cdot t \cdot q$. This reduces the number of trees by $1$.

Otherwise, all occurrences of $X$ are inside $t$. By definition of $\beta_X$, this also includes a subtree for a rule $Y \to u$ with $\abs{u}_X > 0$ and $y_X = y_Y + 1$.  If the non-terminal $Y$ of this subtree is on the path from the root $X$ to the leaf $X$, then we can factorize $t = p \cdot q$ at $Y$ and replace $t$ by $q \cdot p$.
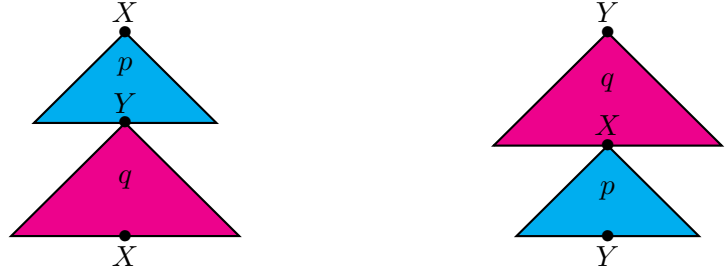
\begin{center}
\hspace*{\fill}
\begin{tikzpicture}[scale=0.3]
\draw[thick,fill=cyan] (0,4) -- (4,0) -- (-4,0) -- cycle;
\draw (0,2.5) node {$p$};
\draw[thick,fill=magenta] (0,0) -- (5,-5) -- (-5,-5) -- cycle;
\draw (0,-2.5) node {$q$};
\draw (0,4) node {$\bullet$} node[above] {$X$};
\draw (0,0) node {$\bullet$} node[above] {$Y$};
\draw (0,-5) node {$\bullet$} node[below] {$X$};
\end{tikzpicture}
\hspace*{\fill}
\begin{tikzpicture}[scale=0.3]
\draw[thick,fill=magenta] (0,5) -- (5,0) -- (-5,0) -- cycle;
\draw (0,2.8) node {$q$};
\draw[thick,fill=cyan] (0,0) -- (4,-4) -- (-4,-4) -- cycle;
\draw (0,-2) node {$p$};
\draw (0,5) node {$\bullet$} node[above] {$Y$};
\draw (0,0) node {$\bullet$} node[above] {$X$};
\draw (0,-4) node {$\bullet$} node[below] {$Y$};
\end{tikzpicture}
\hspace*{\fill}
\end{center}
Otherwise, we can factorize $t = p \cdot q$ at this particular occurrence of $Y$. Since $u$ defines the child nodes of the root $Y$ in $q$ and since $\abs{u}_X > 0$, we can factorize $q = r \cdot r'$ at $X$. Now, we can replace $t$ by $r \cdot p \cdot r'$.
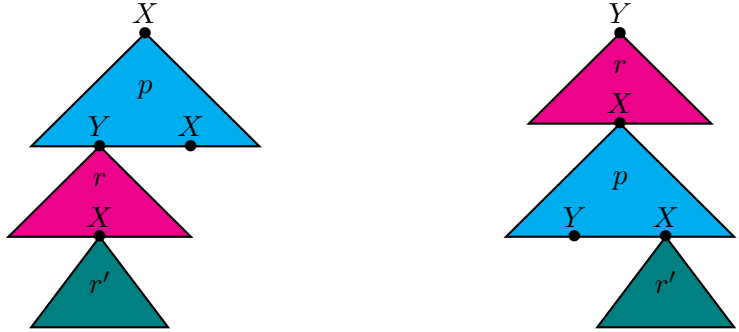
\begin{center}
\hspace*{\fill}
\begin{tikzpicture}[scale=0.3]
\draw[thick,fill=cyan] (0,5) -- (5,0) -- (-5,0) -- cycle;
\draw (0,2.5) node {$p$};
\draw[thick,fill=magenta] (-2,0) -- (2,-4) -- (-6,-4) -- cycle;
\draw (-2,-1.5) node {$r$};
\draw[thick,fill=teal] (-2,-4) -- (1,-8) -- (-5,-8) -- cycle;
\draw (-2,-6) node {$r'$};
\draw (0,5) node {$\bullet$} node[above] {$X$};
\draw (2,0) node {$\bullet$} node[above] {$X$};
\draw (-2,0) node {$\bullet$} node[above] {$Y$};
\draw (-2,-4) node {$\bullet$} node[above] {$X$};
\end{tikzpicture}
\hspace*{\fill}
\begin{tikzpicture}[scale=0.3]
\draw[thick,fill=magenta] (0,4) -- (4,0) -- (-4,0) -- cycle;
\draw (0,2.5) node {$r$};
\draw[thick,fill=cyan] (0,0) -- (5,-5) -- (-5,-5) -- cycle;
\draw (0,-2.5) node {$p$};
\draw[thick,fill=teal] (2,-5) -- (5,-9) -- (-1,-9) -- cycle;
\draw (2,-7) node {$r'$};
\draw (0,4) node {$\bullet$} node[above] {$Y$};
\draw (0,0) node {$\bullet$} node[above] {$X$};
\draw (2,-5) node {$\bullet$} node[above] {$X$};
\draw (-2,-5) node {$\bullet$} node[above] {$Y$};
\end{tikzpicture}
\hspace*{\fill}
\end{center}
In any case, this reduces the index of this root by $1$.
By repeating the previous steps as long as possible, eventually there is only one tree left. This is a derivation tree for a word $w$ with $\abs{w}_a = z_a$ for all $a \in \Sigma$.

The formula $\varphi$ can be computed in linear time since for instance in $\sum_{p \in P} X(p) \cdot x_p$ and $\sum_{p \in P} \abs{p}_X \cdot x_p$ it suffices to consider those rules $p$ with $X(p) \neq 0$ and $\abs{p}_X \neq 0$, respectively.
\qed
\end{proof}

\begin{corollary}[Parikh~\cite{parikh1961,parikh66}]\label{cor:ParikhSL}
  If $L \subseteq \Sigma^*$ is context-free, then $\Psi(L) \subseteq \N^{\abs{\Sigma}}$ is semilinear.
\end{corollary}

\begin{proof}
  If $L$ is context-free, then $\Psi(L)$ is definable in Presburger arithmetic by Theorem~\ref{thm:Parikh}. Ginsburg and Spanier~\cite{GinsburgSpanier1966pjm} showed that Presburger definable subsets of $\N^k$ are semilinear.
  \qed
\end{proof}

%\bibliographystyle{abbrv}
%\bibliography{traces}

\newcommand{\Ju}{Ju}\newcommand{\Ph}{Ph}\newcommand{\Th}{Th}\newcommand{\Ch}{Ch}\newcommand{\Yu}{Yu}\newcommand{\Zh}{Zh}\newcommand{\St}{St}\newcommand{\curlybraces}[1]{\{#1\}}

\end{document}